\documentclass[aps,showpacs,twocolumn]{revtex4}

\usepackage{graphicx}
\usepackage{epstopdf}
\usepackage{dcolumn}
\usepackage[ulem=normalem]{changes}
\usepackage{amsmath}

\begin{document}
\title{Partial dynamical symmetry in Bose-Fermi systems}
\author{P.~Van~Isacker}
\affiliation{Grand Acc\'el\'erateur National d'Ions Lourds,
CEA/DSM--CNRS/IN2P3, B.P.~55027, F-14076 Caen Cedex 5, France}

\author{J.~Jolie}
\affiliation{Institute of Nuclear Physics, University of Cologne,
Z\"ulpicherstrasse 77, 50937 Cologne, Germany}

\author{T.~Thomas}
\affiliation{Institute of Nuclear Physics, University of Cologne,
Z\"ulpicherstrasse 77, 50937 Cologne, Germany}

\author{A.~Leviatan}
\affiliation{Racah Institute of Physics, The Hebrew University, 
Jerusalem 91904, Israel}
\date{\today}

\begin{abstract}
We generalize the notion of partial dynamical symmetry (PDS) to a 
system of interacting bosons and fermions. 
In a PDS, selected states of the Hamiltonian
are solvable and preserve the symmetry exactly,
while other states are mixed. 
As a first example of such novel symmetry construction, 
spectral features of the odd-mass nucleus $^{195}$Pt are analyzed.
\end{abstract}
\pacs{21.60.Fw, 21.10.Re, 21.60.Ev, 27.80+w}
\maketitle

During the last several decades, the concept of dynamical symmetry (DS) 
has become the cornerstone of algebraic modeling of dynamical systems.
It has been applied in many branches of physics,
such as hadronic~\cite{BNB}, nuclear~\cite{ibm,ibfm}, 
atomic~\cite{nano} and molecular physics~\cite{vibron,Frank94}. 
Its basic paradigm is to write the Hamiltonian of the system
in terms of Casimir operators of a chain of nested algebras,
$G_{\rm dyn} \supset G \supset \dots \supset G_{\rm sym}$, 
where $G_{\rm dyn}$ is the dynamical algebra,
in terms of which any model operator of a physical observable 
can be expressed, and $G_{\rm sym}$ is the symmetry algebra. 
A given DS defines a class of many-body Hamiltonians
that admit an analytic solution for all states,
with closed expressions for the energy eigenvalues,
quantum numbers for classification
and definite selection rules for transition processes.

An exact DS provides considerable insights into complex dynamics
and its merits are self evident.
However, in most applications to realistic systems,
its predictions are rarely fulfilled and one is compelled to break it.
The DS spectrum imposes constraints on the pattern of level-splitting
which many times is at variance with the empirical data. 
More often one finds that the assumed symmetry is not obeyed uniformly,
{\it i.e.}, is fulfilled by some of the states but not by others.
The required symmetry breaking is achieved
by including in the Hamiltonian
terms associated with different subalgebra chains of $G_{\rm dyn}$,
resulting in a loss of solvability and pronounced mixing.  
The need to address such situations, but still preserve important symmetry remnants,
has led to the introduction of partial dynamical  symmetry (PDS)~\cite{Alhassid92,Leviatan96}. 
The essential idea is to relax the stringent conditions of {\it complete} solvability
so that only part of the eigenspectrum retains 
analyticity and/or good quantum numbers, in the spirit of quasi-solvable models~\cite{QSM}.
Various types of PDSs were proposed~\cite{Leviatan96,Leviatan86,Isacker99,Leviatan02,Leviatan11} 
and algorithms for constructing Hamiltonians with such property 
have been developed~\cite{Alhassid92,GarciaRamos09}. 
Bosonic Hamiltonians with PDS
have been applied to nuclear 
spectroscopy~\cite{Leviatan96,Leviatan86,Isacker99,Leviatan02,Leviatan11,
GarciaRamos09,Leviatan13,Casten14,Kremer14}, 
where extensive tests provide empirical evidence for their relevance to a broad range of nuclei. 
Similar PDS Hamiltonians have been used in molecular spectroscopy~\cite{Ping97}
and in the study of quantum phase transitions~\cite{Leviatan07,Macek14} 
and of mixed regular and chaotic dynamics~\cite{Macek14,WAL93}.
Fermionic Hamiltonians with PDS have been identified within the nuclear 
shell model and applied to light nuclei~\cite{Escher00}
and seniority isomers~\cite{Rowe01,Isacker08}. 
The growing number of empirical manifestations 
suggests a more pervasive role of PDSs in dynamical systems
than heretofore realized.

All examples of PDS considered so far,
were confined to systems of a given statistics (bosons or fermions). 
In this Rapid Communication,
we extend the PDS concept to mixed systems of bosons and fermions, 
and present an empirical example of this novel construction. 
Systems with such composition of constituents are of broad interest
and arise, for example, in the study of rotation-vibration-electronic 
spectra in molecules, collective states in odd-mass nuclei, 
electron-phonon phenomena in crystals,
and spin-boson models in quantum optics.

If the separate numbers of bosons $N$ and fermions $M$ are conserved,
the dynamical algebra of a Bose-Fermi system is of product form
\begin{equation}
\begin{array}{ccc}
{\rm U}^{\rm B}(\Omega_{\rm B})&\otimes&{\rm U}^{\rm F}(\Omega_{\rm F})\\
\downarrow&&\downarrow\\[0mm]
[N]&&[1^M]
\end{array},
\label{e_dynalg}
\end{equation}
where $\Omega_{\rm B}$ ($\Omega_{\rm F}$)
is the number of states available to a single boson (fermion).
The statistics among the particles is imposed
by an appropriate choice of irreducible representation (irrep), 
symmetric and anti-symmetric, for the bosons and fermions, respectively,
as indicated in Eq.~(\ref{e_dynalg}).
There exist several strategies to define 
DSs with ${\rm U}^{\rm B}(\Omega_{\rm B})\otimes{\rm U}^{\rm F}(\Omega_{\rm F})$
as a starting point~\cite{ibfm}.
They all define a chain of nested subalgebras,
relying on the existence of isomorphisms between boson and fermion algebras
and ending in the symmetry algebra.

Let us for the sake of concreteness
consider a particular example
while emphasizing that results of this Rapid Communication are of a generic nature
that apply to any quantum-mechanical problem of interacting bosons 
and fermions, as long as it can be 
formulated in an algebraic language.
We consider $N$ bosons with angular momentum $\ell=0$ ($s$) or $\ell=2$ ($d$) 
coupled to a single ($M=1$) fermion with 
angular momentum $j=1/2$, 3/2, or 5/2.
This corresponds to the choice $\Omega_{\rm B}=6$ and $\Omega_{\rm F}=12$,
and a possible classification is as follows:
\begin{widetext}
\begin{equation}
\begin{array}{ccccc ccccccccc c}
{\rm U}^{\rm B}(6)&\otimes&{\rm U}^{\rm F}(12)&\supset&
\Big({\rm U}^{\rm BF}(6)&\supset&{\rm SO}^{\rm BF}(6)&
\supset&{\rm SO}^{\rm BF}(5)&\supset&{\rm SO}^{\rm BF}(3)\Big)
&\otimes&{\rm SU}^{\rm F}(2)&\supset&{\rm Spin}^{\rm BF}(3)\\
\downarrow&&\downarrow&&\downarrow&&\downarrow&&\downarrow&&\downarrow&&
\downarrow&&\downarrow\\[0mm]
[N]&&[1^M]&&[N_1,N_2]&&\langle\sigma_1,\sigma_2\rangle&&
(\tau_1,\tau_2)&&L&&\tilde{s}&&J
\end{array},
\label{e_clas}
\end{equation}
\end{widetext}
where underneath each algebra the associated irrep labels are 
indicated, and $G^{\rm BF}$ is the direct sum of 
$G^{\rm B}$ and $G^{\rm F}$. 
For $M=0$, the classification~(\ref{e_clas}) reduces to
the SO(6) limit of the interacting boson model~\cite{Arima79}
which is of relevance for the even-even platinum isotopes~\cite{Cizewski78}. 
For $M=1$, the classification~(\ref{e_clas}) is proposed
in the context of the interacting boson-fermion model (IBFM)~\cite{ibfm} 
to describe odd-mass isotopes of platinum
with the odd neutron in the orbits $3p_{1/2}$, $3p_{3/2}$, and $2f_{5/2}$,
which are dominant for these isotopes~\cite{Isacker84,Bijker85}.
Since we are interested here in Bose-Fermi systems,
we apply the classification~(\ref{e_clas}) for $M=1$, which 
implies $\tilde{s}=1/2$, and refer to it as the ${\rm SO}^{\rm BF}(6)$ limit.

The eigenstates~(\ref{e_clas}) are obtained with a Hamiltonian
that is a combination of Casimir operators $\hat C_n[G]$
of order $n$ of an algebra $G$ appearing in the chain. 
Up to a constant energy, this Hamiltonian is of the form
\begin{eqnarray}
\hat H_{\rm DS}&=&
a\,\hat C_2[{\rm U}^{\rm BF}(6)]+
b\,\hat C_2[{\rm SO}^{\rm BF}(6)]+
c\,\hat C_2[{\rm SO}^{\rm BF}(5)]
\nonumber\\&&+
d\,\hat C_2[{\rm SO}^{\rm BF}(3)]+
d'\hat C_2[{\rm Spin}^{\rm BF}(3)].
\label{e_hamds}
\end{eqnarray}
The associated eigenvalue problem is analytically solvable,
leading to the energy expression
\begin{eqnarray}
E_{\rm DS}&=&
a\,f_5(N_1,N_2)+
b\,f_4(\sigma_1,\sigma_2)+
c\,f_3(\tau_1,\tau_2)
\nonumber\\&&+
d\,L(L+1)+
d'J(J+1),
\label{e_eval}
\end{eqnarray}
with $f_i(s_1,s_2)\equiv s_1(s_1+i)+s_2(s_2+i-2)$.
The energy spectrum of the Hamiltonian~(\ref{e_hamds})
is then determined once the allowed values of
$[N_1,N_2]$, $\langle\sigma_1,\sigma_2\rangle$, $(\tau_1,\tau_2)$, $L$, 
and $J$ for a given $N$ and $M=1$ are found.
Such branching rules can be obtained
with standard group-theoretical techniques~\cite{Wybourne74}.
\begin{table*}
\caption{\label{t_tensors}
Two-particle tensor operators in the ${\rm SO}^{\rm BF}(6)$ limit.
The superscript ${\cal L}({\cal J})$ stands for the coupling 
${\cal J}={\cal L}\pm 1/2$.}
\begin{ruledtabular}
\begin{tabular}{ccccccccl}
$N$&$M$&$[N_1,N_2]$&$\langle\sigma_1,\sigma_2\rangle$&$(\tau_1,\tau_2)$&
${\cal L}$&${\cal J}$&&
Tensor operator ${\cal T}^{{\cal L}({\cal J})}_{+,M_{\cal J}}$\\[2pt]
\hline
$2$&$0$&$[2,0]$&$\langle0,0\rangle$&$(0,0)$&$0$&$0$&&
${\cal V}^{0(0)}_+\equiv
\sqrt{\frac5{12}}(d^\dag d^\dag)^{(0)}_0
-\sqrt{\frac1{12}}(s^\dag s^\dag)^{(0)}_0$\\[2pt]
$1$&$1$&$[2,0]$&$\langle0,0\rangle$&$(0,0)$&$0$&$1/2$&&
${\cal V}^{0(1/2)}_{+,\mu}\equiv
-\sqrt{\frac16}(s^\dag a^\dag_{1/2})^{(1/2)}_\mu-
\sqrt{\frac13}(d^\dag a^\dag_{3/2})^{(1/2)}_\mu+
\sqrt{\frac12}(d^\dag a^\dag_{5/2})^{(1/2)}_\mu$\\[2pt]
$1$&$1$&$[1,1]$&$\langle1,1\rangle$&$(1,1)$&$1$&$1/2$&&
${\cal U}^{1(1/2)}_{+,\mu}\equiv
\sqrt{\frac35}(d^\dag a^\dag_{3/2})^{(1/2)}_\mu+
\sqrt{\frac25}(d^\dag a^\dag_{5/2})^{(1/2)}_\mu$\\[2pt]
$1$&$1$&$[1,1]$&$\langle1,1\rangle$&$(1,1)$&$1$&$3/2$&&
${\cal U}^{1(3/2)}_{+,\mu}\equiv
-\sqrt{\frac{3}{10}}(d^\dag a^\dag_{3/2})^{(3/2)}_\mu+
\sqrt{\frac{7}{10}}(d^\dag a^\dag_{5/2})^{(3/2)}_\mu$\\[2pt]
$1$&$1$&$[1,1]$&$\langle1,1\rangle$&$(1,0)$&$2$&$3/2$&&
${\cal U}^{2(3/2)}_{+,\mu}\equiv
\sqrt{\frac12}(s^\dag a^\dag_{3/2})^{(3/2)}_\mu-
\sqrt{\frac12}(d^\dag a^\dag_{1/2})^{(3/2)}_\mu$\\[2pt]
$1$&$1$&$[1,1]$&$\langle1,1\rangle$&$(1,0)$&$2$&$5/2$&&
${\cal U}^{2(5/2)}_{+,\mu}\equiv
\sqrt{\frac12}(s^\dag a^\dag_{5/2})^{(5/2)}_\mu-
\sqrt{\frac12}(d^\dag a^\dag_{1/2})^{(5/2)}_\mu$\\[2pt]
$1$&$1$&$[1,1]$&$\langle1,1\rangle$&$(1,1)$&$3$&$5/2$&&
${\cal U}^{3(5/2)}_{+,\mu}\equiv
\sqrt{\frac45}(d^\dag a^\dag_{3/2})^{(5/2)}_\mu+
\sqrt{\frac15}(d^\dag a^\dag_{5/2})^{(5/2)}_\mu$\\[2pt]
$1$&$1$&$[1,1]$&$\langle1,1\rangle$&$(1,1)$&$3$&$7/2$&&
${\cal U}^{3(7/2)}_{+,\mu}\equiv
-\sqrt{\frac{1}{10}}(d^\dag a^\dag_{3/2})^{(7/2)}_\mu+
\sqrt{\frac{9}{10}}(d^\dag a^\dag_{5/2})^{(7/2)}_\mu$\\[2pt]
\end{tabular}
\end{ruledtabular}
\end{table*}
While $\hat{H}_{\rm DS}$~(\ref{e_hamds}) is {\em completely} solvable,
the question arises whether terms can be added
that preserve solvability for {\em part} of its spectrum.
This can be achieved by the construction of a PDS.

The algorithm to construct a PDS~\cite{GarciaRamos09} 
starts from the character under the classification~(\ref{e_clas})
of the boson and fermion creation operators
$b^\dag_{\ell m_\ell}$ and $a^\dag_{jm_j}$~\cite{ibfm}.
Annihilation operators $b_{\ell m_\ell}$ and $a_{jm_j}$ 
transform in the same manner under orthogonal algebras
if they are modified according to
$\tilde b_{\ell m_\ell}\equiv(-)^{\ell+m_\ell}b_{\ell,-m_\ell}$
and $\tilde a_{jm_j}\equiv(-)^{j+m_j}a_{j,-m_j}$.
The single-fermion angular momentum ($j=1/2,3/2,5/2$) 
can be divided into a pseudo-orbital angular momentum $(\tilde\ell\!=\!0,2)$ 
coupled to a pseudo-spin ($\tilde s\!=\!1/2$). 
The resulting $\tilde\ell$-$\tilde s$ basis is given by
$c^{\dag}_{\tilde\ell\tilde m_\ell;\tilde s\tilde m_s}=
\sum_{j,m}(\tilde\ell,\tilde m_{\ell};\tilde s,\tilde m_s\vert j,m)\,a^{\dagger}_{jm}$. 

Composite operators with definite tensor character under the 
classification~(\ref{e_clas}) 
can be constructed by use of generalized coupling coefficients
which can be written as a product of
${\rm U}(6)\supset{\rm SO}(6)$,
${\rm SO}(6)\supset{\rm SO}(5)$,
and ${\rm SO}(5)\supset{\rm SO}(3)$ isoscalar factors~\cite{Wybourne74}.
For the two-particle operators
(needed for the construction of a two-body interaction)
the tensor character is uniquely specified by the SO(6) and SO(5) labels
$\langle\sigma_1,\sigma_2\rangle$ and $(\tau_1,\tau_2)$,
together with the SO$^{\rm BF}$(3) and Spin$^{\rm BF}$(3) labels 
$\cal L$ and $\cal J$.
For example, operators that create a boson and a fermion
with tensor character
$[N_1,N_2]$ $\langle\sigma_1,\sigma_2\rangle$ $(\tau_1,\tau_2)$ ${\cal LJ}$ 
are $$\sum_{\tau\tilde\tau}
\left\langle\begin{array}{cc|c}
\langle1\rangle&\langle1\rangle&\langle\sigma_1,\sigma_2\rangle\\
(\tau)&(\tilde\tau)&(\tau_1,\tau_2)
\end{array}
\right\rangle
\sum_j\bar{\cal L}\bar\jmath
\Bigl\{\begin{array}{ccc}
\ell&\tilde\ell&{\cal L}\\
\tilde s&{\cal J}&j
\end{array}\Bigr\}
(b^\dag_\ell a^\dag_j)^{({\cal J})}_{M_{\cal J}},$$
with $(\ell,\tilde\ell)=0,2$, for $(\tau,\tilde\tau)=0,1$, respectively, 
$j\!=\!\tilde\ell\pm 1/2$, $\bar x\equiv\sqrt{2x+1}$ 
and the symbol $\left\langle:\,:|:\right\rangle$ 
is an ${\rm SO}(6)\supset{\rm SO}(5)$ isoscalar factor, 
tabulated in Ref.~\cite{ibfm}.
Expressions for the two-particle creation operators of interest here
are given in Table~\ref{t_tensors}.
The corresponding annihilation operators with the correct tensor 
properties follow from
$\tilde{\cal T}^{{\cal L}({\cal J})}_{-,M_{\cal J}} \!=\!
(-)^{{\cal J}+M_{\cal J}}\left({\cal T}^{{\cal L}({\cal J})}_{+,-M_{\cal J}}\right)^\dag$, 
where ${\cal T}\!=\!{\cal U}$ or $\cal V$.

The lowest-lying states in the spectrum of an odd-mass nucleus,
described in terms of $N$ bosons and one fermion,
can be written as $|[N+1]\langle N+1\rangle(\tau)LJM_J\rangle$;
the next class of states belongs to
$|[N,1]\langle N,1\rangle(\tau_1,\tau_2)LJM_J\rangle$
while there is also some evidence from one-neutron transfer 
for $|[N,1]\langle N-1\rangle(\tau)LJM_J\rangle$ states~\cite{Metz00}. 
All two-particle operators listed in Table~\ref{t_tensors}
annihilate particular states, hence lead to a PDS of some kind.
For example, the operators
with ${\rm U}^{\rm BF}(6)$ labels $[N_1,N_2]=[1,1]$ satisfy 
\begin{equation}
\tilde{\cal U}^{{\cal L}({\cal J})}_{-,M_{\cal J}}
|[N+1]\langle\sigma\rangle(\tau)LJM_J\rangle=0,
\label{e_anni1}
\end{equation}
for all permissible $(\sigma \tau L J M_J)$. 
This is so because a state with $N-1$ bosons and no fermion
has the ${\rm U}^{\rm BF}(6)$ label $[N-1]$.
Given the multiplication $[N-1]\otimes[1,1]=[N,1]\oplus[N-1,1,1]$, 
the action of a ${\cal U}^{{\cal L}({\cal J})}_{+,-M_{\cal J}}$ operator
on an $(N-1)$-boson state can never yield a boson-fermion state
with the ${\rm U}^{\rm BF}(6)$ labels $[N+1]$.
Similar arguments involving SO(6) multiplication
lead to the following properties for the $\cal V$ operators
which have SO(6) tensor character $\langle0,0\rangle$:
\begin{subequations}
\begin{eqnarray}
&&\tilde{\cal V}^{0({\cal J})}_{-,M_{\cal J}}
|[N+1]\langle N+1\rangle(\tau)LJM_J\rangle=0,\\
&&\tilde{\cal V}^{0({\cal J})}_{-,M_{\cal J}}
|[N,1]\langle N,1\rangle(\tau_1,\tau_2)LJM_J\rangle=0.
\end{eqnarray}
\label{e_anni2}
\end{subequations}

An alternative way of constructing Hamiltonians with PDS for an algebra $G$,
is to identify $n$-particle operators which annihilate a lowest-weight 
state of a prescribed $G$-irrep~\cite{Alhassid92}. 
In the IBFM, such a state, which transforms as $[N+1]$ and $\tilde s=1/2$
under ${\rm U}^{\rm BF}(6)\otimes {\rm SU}^{\rm F}(2)$, is given~by
\begin{equation}
\vert \Psi_{\rm g}\rangle \propto  
[b^{\dag}_c(\beta)]^Nf^{\dag}_{\tilde m_s}(\beta)\vert 0\rangle,
\label{Psi-g}
\end{equation}
where $b^{\dag}_{c}(\beta)  \propto (\beta\, d^{\dag}_0 + s^{\dag})$ 
and $f^{\dag}_{\tilde m_s}(\beta) \propto 
(\beta\, c^{\dag}_{2,0;1/2,\tilde m_s} + c^{\dag}_{0,0;1/2,\tilde m_s})$ 
in the $\tilde\ell$-$\tilde s$ basis defined above. 
$\vert \Psi_{\rm g}\rangle$ is a condensate of $N$ bosons and a single fermion, 
and represents an intrinsic state for the ground band with deformation $\beta$.
The Hermitian conjugate of the following 
two-particle operators
\begin{subequations}
\begin{eqnarray}
{\cal V}^{0(0)}_+ &=&
{\textstyle\sqrt{\frac{5}{12}}}(d^\dag d^\dag)^{(0)}_0
-{\textstyle\frac{\beta^2}{\sqrt{12}}}(s^\dag s^\dag)^{(0)}_0,
\label{Tb1}\\
{\cal V}^{0(1/2)}_{+,\mu} &=&
\sqrt{\tfrac{5}{6}}(d^{\dag} c^{\dag}_{2;1/2})^{0(1/2)}_{\mu}
-\tfrac{\beta^2}{\sqrt{6}}(s^{\dag} c^{\dag}_{0;1/2})^{0(1/2)}_{\mu},\qquad
\label{Tb2}\\
{\cal U}^{{\cal L}({\cal J})}_{+,\mu} &=&
(d^{\dag} c^{\dag}_{2;1/2})^{{\cal L}({\cal J})}_{\mu},\quad
{\textstyle{\cal L}=1,3},
\label{Tb3}\\
{\cal U}^{2({\cal J})}_{+,\mu} &=&
(s^{\dag} c^{\dag}_{2;1/2})^{2({\cal J})}_{\mu}
- (d^{\dag} c^{\dag}_{0;1/2})^{2({\cal J})}_{\mu},
\label{Tb4}
\end{eqnarray}
\label{Tb}
\end{subequations}
satisfy $\tilde{\cal T}^{{\cal L}({\cal J})}_{-,\mu}\vert \Psi_{\rm g}\rangle=0$. 
The $\cal V$ operators of Eqs.~(\ref{Tb1})-(\ref{Tb2}) 
satisfy also $\tilde{\cal V}^{0({\cal J})}_{-,\mu}\vert \Psi_{\rm e}\rangle=0$, 
where 
\begin{equation}
\vert \Psi_{\rm e}\rangle \propto  
[b^{\dag}_c(\beta)\,c^{\dag}_{2,1;1/2,\tilde m_s} 
- d^{\dag}_{1}\,f^{\dag}_{\tilde m_s}(\beta)][b^{\dag}_c(\beta)]^{N-1}
\vert 0\rangle\;
\label{Psi-e}
\end{equation}
is an intrinsic state, with ${\rm U}^{\rm BF}(6)$ label $[N,1]$, 
representing an excited band in the odd-mass nucleus.
For $\beta=1$, $\vert \Psi_{\rm g}\rangle$ and $\vert \Psi_{\rm e}\rangle$ 
become the lowest-weight states in the ${\rm SO}^{\rm BF}(6)$ irreps 
$\langle N+1\rangle$ and $\langle N,1\rangle$, respectively, 
from which the $|(\tau_1,\tau_2)LJM_J\rangle$ states of Eq.~(\ref{e_anni2}) 
can be obtained by ${\rm SO}^{\rm BF}(5)$ projection,
and the operators~(\ref{Tb}) coincide with those listed in 
Table~\ref{t_tensors}.

The combined effect of normal-ordered interactions
constructed out of the $\cal T$ operators in Table~\ref{t_tensors}
added to the DS Hamiltonian~(\ref{e_hamds}),
gives rise to a rich variety of possible PDSs.
In the current application to $^{195}$Pt
we take a restricted Hamiltonian of the form
\begin{eqnarray}
\hat H_{\rm PDS} &=&
\hat H_{\rm DS}+
a_0\hat V_0^{1/2}+
a'_1(2\hat U_1^{1/2}-\hat U_1^{3/2})+
\nonumber\\
&&
a_2(\hat U_2^{3/2}+\hat U_2^{5/2})+
a_3(\hat U_3^{5/2}+\hat U_3^{7/2}),\qquad
\label{e_hampds}
\end{eqnarray}
in terms of the interactions
$\hat T_{\cal L}^{\cal J}\equiv\bar{\cal J}({\cal T}^{\cal L(J)}_+
\tilde{\cal T}^{\cal L(J)}_-)^{(0)}_0$, where 
${\cal T}={\cal U}$ or $\cal V$ and $T=U$ or $V$. 
These interactions can be transcribed 
as tensors with total pseudo-orbital $\tilde L$ and pseudo-spin $\tilde S$
coupled to zero total angular momentum.
In particular, the $a'_1$ term in Eq.~(\ref{e_hampds}) has 
$\tilde L\!=\!\tilde S\!=\!1$, 
while the $a_0,\, a_2$ and $a_3$ terms have $\tilde L\!=\!\tilde S\!=\!0$.
\begin{figure*}
\centering
\includegraphics[width=10cm]{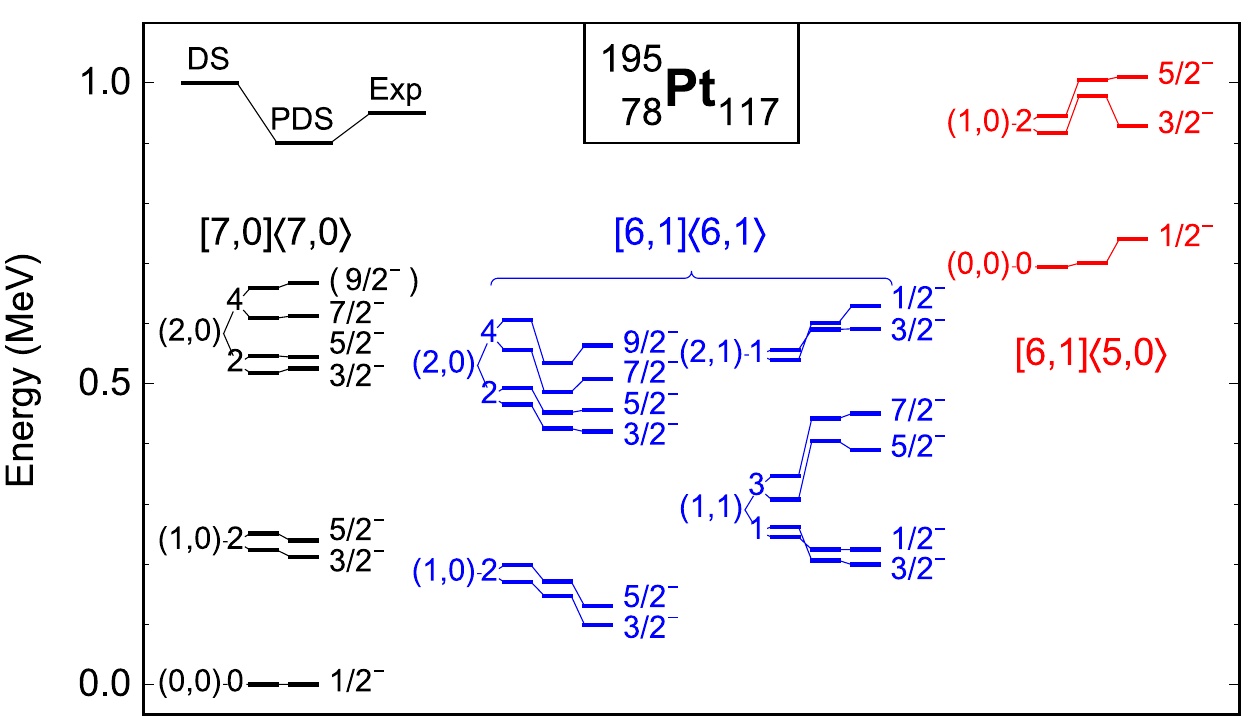}
\includegraphics[width=7cm]{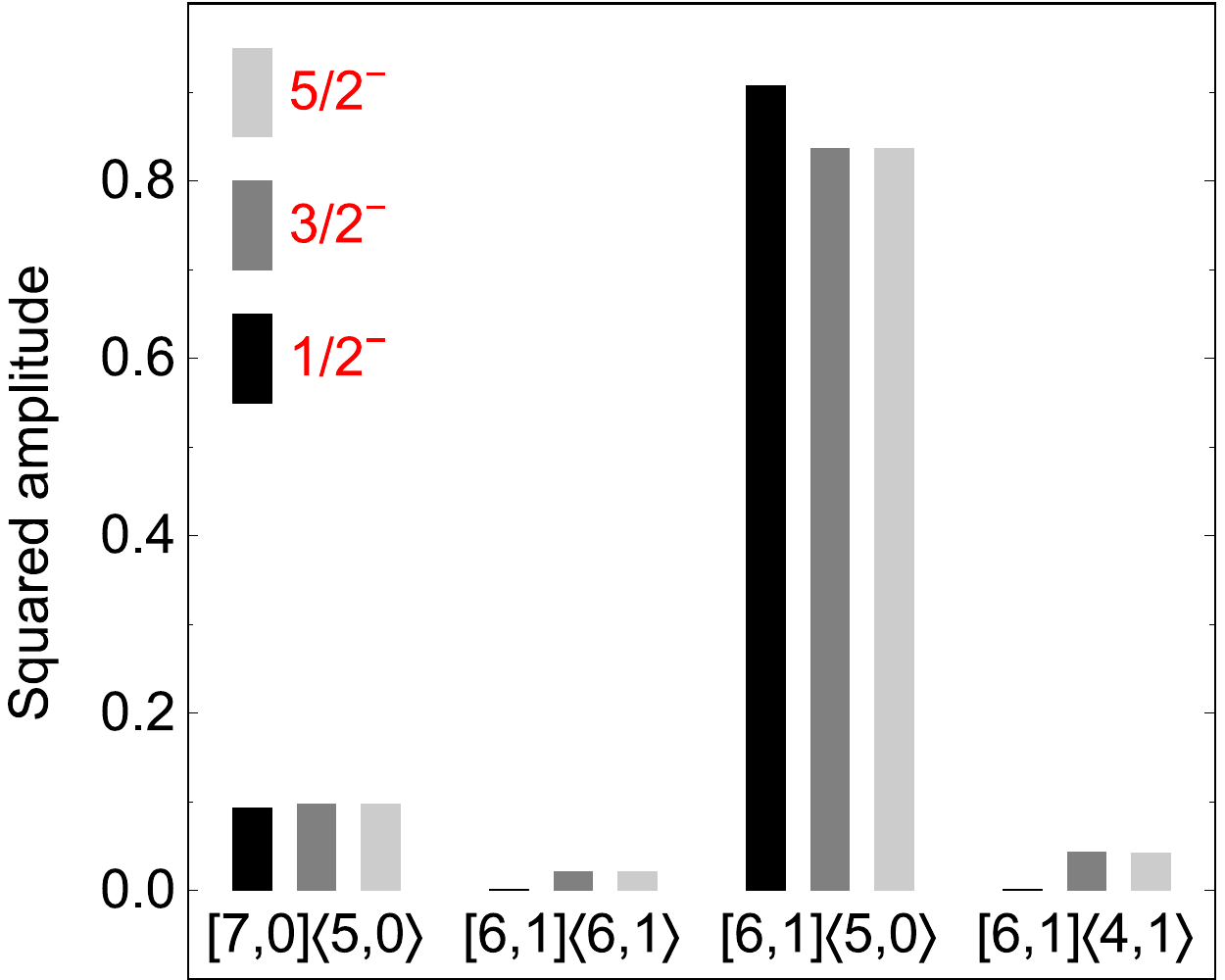}
\caption{(Color online). Left panel: 
Observed and calculated energy spectrum of $^{195}$Pt. 
The levels in black are the solvable $[7,0]\langle7,0\rangle$
eigenstates of $\hat H_{\rm DS}$~(\ref{e_hamds}),
whose structure and energy remain unaffected by the added PDS interactions 
in Eq.~(\ref{e_hampds}). 
The levels in blue (red) are the 
$[6,1]\langle6,1\rangle$ ($[6,1]\langle5,0\rangle$) eigenstates 
of $\hat H_{\rm DS}$~(\ref{e_hamds})
and are subsequently perturbed by the PDS interactions in Eq.~(\ref{e_hampds}).
Right panel: ${\rm SO}^{\rm BF}(6)$ decomposition of the eigenstates of 
$\hat H_{\rm PDS}$~(\ref{e_hampds}), shown in red on the left panel.}
\label{f_pt195e}
\end{figure*}
\begin{table}
\caption{Observed 
$B({\rm E}2;J_{\rm i}\rightarrow J_{\rm f})$ values
between negative-parity states in $^{195}$Pt
compared with the DS and PDS predictions of the ${\rm SO}^{\rm BF}(6)$ limit.
The solvable (mixed) states are members of the ground (excited) bands 
shown in Fig.~\ref{f_pt195e}.
The E2 operator employed is defined in the text.} 
\label{t_be2}
\begin{ruledtabular}
\begin{tabular}{rrccrrccrrrrrr}
\noalign{\smallskip}
${E_{\rm i}}$&&$J_{\rm i}$&~~~&
${E_{\rm f}}$&&$J_{\rm f}$&~~~~~~&
\multicolumn{5}{c}
{$B({\rm E}2;J_{\rm i}\rightarrow J_{\rm f}$) ($10^{-3}~e^2{\rm b}^2$)}\\
\cline{9-13}
(keV)&&&&(keV)&&&&Exp&~~~~~~&DS&~~~~~~&PDS\\
\hline\noalign{\smallskip}
&\multicolumn{12}{c}{Solvable $\rightarrow$ solvable}&\\[1pt]
212&&${3/2}$&&0&&${1/2}$&&$190(10)$&&179&&179\\
239&&${5/2}$&&0&&${1/2}$&&$170(10)$&&179&&179\\
525&&${3/2}$&&0&&${1/2}$&&$17(1)$&&0&&0\\
525&&${3/2}$&&239&&${5/2}$&&$\le19$&&72&&72\\
544&&${5/2}$&&0&&${1/2}$&&$8(4)$&&0&&0\\
612&&${7/2}$&&212&&${3/2}$&&$170(70)$&&215&&215\\
667&&${9/2}$&&239&&${5/2}$&&$200(40)$&&239&&239\\[2pt]
&\multicolumn{12}{c}{Solvable $\rightarrow$ mixed}&\\[1pt]
239&&${5/2}$&&99&&${3/2}$&&$60(20)$&&0&&0\\
525&&${3/2}$&&99&&${3/2}$&&$\le33$&&7&&3\\
525&&${3/2}$&&130&&${5/2}$&&$9(5)$&&3&&2\\
612&&${7/2}$&&99&&${3/2}$&&$5(3)$&&9&&11\\
667&&${9/2}$&&130&&${5/2}$&&$12(3)$&&10&&12\\[2pt]
&\multicolumn{12}{c}{Mixed $\rightarrow$ solvable}&\\[1pt]
99&&${3/2}$&&0&&${1/2}$&&$38(6)$&&35&&34\\
130&&${5/2}$&&0&&${1/2}$&&$66(4)$&&35&&33\\
420&&${3/2}$&&0&&${1/2}$&&$15(1)$&&0&&0\\
456&&${5/2}$&&0&&${1/2}$&&$\le0.04$&&0&&0\\
508&&${7/2}$&&212&&${3/2}$&&$55(17)$&&20&&18\\
563&&${9/2}$&&239&&${5/2}$&&$91(22)$&&22&&22\\
199&&${3/2}$&&0&&${1/2}$&&$25(2)$&&0&&0\\
390&&${5/2}$&&0&&${1/2}$&&$7(1)$&&0&&0\\[2pt]
&\multicolumn{12}{c}{Mixed $\rightarrow$ mixed}&\\[1pt]
420&&${3/2}$&&99&&${3/2}$&&$5(4)$&&177&&165\\
508&&${7/2}$&&99&&${3/2}$&&$240(50)$&&228&&263\\
563&&${9/2}$&&130&&${5/2}$&&$240(40)$&&253&&284\\
390&&${5/2}$&&99&&${3/2}$&&$200(70)$&&219&&179\\
390&&${5/2}$&&130&&${5/2}$&&$\le14$&&55&&35\\
\end{tabular}
\end{ruledtabular}
\end{table}

The experimental spectrum of $^{195}$Pt is shown in Fig.~\ref{f_pt195e}, 
compared with the DS and PDS calculations. The coefficients $c$, $d$, 
and $d'$ in $\hat H_{\rm DS}$~(\ref{e_hamds}) are adjusted to the 
excitation energies of the $[7,0]\langle7,0\rangle$ levels
which are reproduced with a root-mean-square (rms) deviation of 12~keV.
The remaining two coefficients $a$ and $b$ are obtained from an overall fit.
The resulting (DS) values are (in keV):
$a=45.3$, $b=-41.5$, $c=49.1$, $d=1.7$, and $d'=5.6$.
The fit for the PDS calculation proceeds in stages.
First, the parameters $c$, $d$, and $d'$ in Eq.~(\ref{e_hamds}) are taken 
at their DS values. 
This ensures the same spectrum for the $[7,0]\langle7,0\rangle$ levels 
(drawn in black in Fig.~\ref{f_pt195e})
which remain eigenstates of $\hat H_{\rm PDS}$~(\ref{e_hampds}). 
Next, one considers the $[6,1]\langle6,1\rangle$ levels
and improves their description by adding the three PDS $U$ interactions.
The resulting coefficients are (in keV): 
$a'_1=10$, $a_2=-97$, and $a_3=112$.
Eq.~(\ref{e_anni1}) ensures that 
the energies of the $[7,0]\langle7,0\rangle$ levels do not change
while the agreement for the $[6,1]\langle6,1\rangle$ levels is improved
(blue levels in Fig.~\ref{f_pt195e}).
The rms deviation for both classes of levels is 20~keV. 
In particular, unlike in the DS calculation, it is possible to reproduce
the observed inversion of the $1/2^-$-$3/2^-$ doublets
without changing the order of other doublets. 
The additional PDS terms necessitate a readjustment of the $a$ coefficient
in Eq.~(\ref{e_hamds}), for which the final (PDS) value is $a=37.7$~keV,
while the coefficient $b$ is kept unchanged. 
Finally, the position of the $[6,1]\langle5,0\rangle$ levels
(red levels in Fig.~\ref{f_pt195e})
is corrected by considering the PDS $V$ interaction with $a_0=306$~keV 
which, due to Eq.~(\ref{e_anni2}),  
has a marginal effect on lower bands. 
The rms deviation for all levels shown in Fig.~\ref{f_pt195e} is 23~keV. 
While the states $[7,0]\langle7,0\rangle$ of the ground band are pure, 
other eigenstates of $\hat{H}_{\rm PDS}$ in excited bands
can have substantial ${\rm SO}^{\rm BF}(6)$ mixing
(see right panel of Fig.~\ref{f_pt195e}).

A large amount of information also exists
on electromagnetic transition rates and spectroscopic strengths.
In Table~\ref{t_be2}, 25 measured $B$(E2) values in $^{195}$Pt
are compared with the DS and PDS predictions.
The same E2 operator is used as in previous studies~\cite{Bruce85,Mauthofer86}
of the ${\rm SO}^{\rm BF}(6)$ limit,
$\hat{T}_\mu({\rm E2})=e_{\rm b}\hat{Q}^{\rm B}_\mu+e_{\rm f}\hat{Q}^{\rm F}_\mu$,
where $\hat{Q}^{\rm B}_\mu=s^\dag\tilde d_\mu+d^\dag_\mu s$
is the boson quadrupole operator,
$\hat{Q}^{\rm F}_\mu$ is its fermion analogue~\cite{ibfm},
and $e_{\rm b}$ and $e_{\rm f}$ are effective boson and fermion charges,
with the values $e_{\rm b}=-e_{\rm f}=0.151$~$e$b.
Table~\ref{t_be2} is subdivided in four parts
according to whether the initial and/or final state in the transition
has a DS structure (as in Refs.~\cite{Bruce85,Mauthofer86})
or whether it is mixed by the PDS interaction.
It is seen that when both have a DS structure
the $B$(E2) value does not change,
only slight differences occur when either the initial or the final state is mixed,
and the biggest changes arise when both are mixed.

In summary, we have proposed a novel extension of the PDS notion 
to Bose-Fermi systems and exemplified it in $^{195}$Pt.
The analysis highlights the ability of a PDS to select 
and add to the Hamiltonian, in a controlled fashion, 
required symmetry-breaking terms,
yet retain a good DS for a segment of the spectrum.
These virtues greatly enhance the scope of 
applications of algebraic modeling of complex systems. 
The operators~(\ref{Tb}) with $\beta\neq 1$, 
can be used to explore additional PDSs in odd-mass nuclei. 
Partial supersymmetry, of relevance to nuclei~\cite{Metz99}, can be studied
by embedding the algebras of Eq.~({\ref{e_dynalg}}) in a graded Lie algebra.
Work in these directions is in progress.

We thank Stefan Heinze for his help with the numerical calculations. 
J.J. acknowledges financial support by GANIL
and by the Interuniversity Attraction Poles programme
initiated by the Belgian Science Policy Officer
under grant BrixNetwork P7/12 during a sabbatical stay at GANIL.
A.L. acknowledges the hospitality of the Theoretical Division at LANL 
during a sabbatical stay, and the support by the Israel Science Foundation.

\end{document}